# Effects of Filler Configuration and Moisture on Dissipation Factor And Critical Electric Field of Epoxy Composites for HV-ICs Encapsulation

Woojin Ahn[a], Davide Cornigli[b], Dhanoop Varghese[c], Luu Nguyen[d], Srikanth Krishnan[c], Susanna Reggiani[b], and Muhammad Ashraful Alam[a,*]

[a]Department of ECE, Purdue University, West Lafayette, IN USA.
[b]Advanced Research Center for Electronic Systems and Dept. of Electronics, University of Bologna, Bologna, Italy.
[c]Texas Instruments Incorporated, Dallas, TX USA.
[d]Texas Instruments Incorporated, Santa Clara, CA USA.

**Abstract**

Molding compounds (MCs) have been used extensively as an encapsulation material for integrated circuits, however, MCs are susceptible to moisture and charge spreading over time. The increase in dissipation factor due to increase of parasitic electrical conductivity ($\sigma$) and the decrease in dielectric strength ($E_{MC}^{Crit}$) restrict their applications. Thus, a fundamental understanding of moisture transport will suggest strategies to suppress moisture diffusion and broaden their applications. In this paper, we 1) propose a generalized effective medium and solubility (*GEMS*) Langmuir model to quantify water uptake as a function of filler configuration and relative humidity; 2) investigate dominant impact of reacted-water on $\sigma$ through numerical simulations, mass-uptake, and DC conductivity measurements; 3) investigate electric field distribution to explain how moisture ingress reduces $E_{MC}^{Crit}$; and finally 4) optimize the filler configuration to lower the dissipation factor, and enhance $E_{MC}^{Crit}$. The *GEMS*-Langmuir model can be used for any application (e.g., photovoltaics, biosensors) where moisture diffusion leads to reliability challenges.

1. Introduction

Typical plastic mold compounds (MCs) consist of inorganic fillers, such as fused silica or organic clay, embedded within an epoxy resin. The MCs are widely used as a protective encapsulant of ICs, due to their mechanical strength, conformal coverage, simple processing, and low-cost. Sometimes, an IC may be exposed to extreme (hot and humid) environments for a long time, particularly for high power applications in robotics, self-driving car, etc. For these systems, moisture from the external environment and a low level of ionic contamination within the MCs often leads to serious reliability challenges. For example, moisture ingress leads to corrosion, delamination, hygroswelling, and charge spreading on the Back End Of Line (BEOL) region (due to significant injection from high voltage bond wires) parasitically shifts the threshold voltage of the transistors [1]. The charge accumulation at the passivation/mold compound interface is known to directly affect the long-term stability of power devices in high-temperature reverse-bias stress tests, requiring a re-design of the top metal/poly-silicon layers on the active regions [2]. Thus for all these applications, the MCs must be carefully optimized to suppress moisture ingress and charge spreading [3]. Specifically, the MC optimization involves tailoring the inorganic filler size, shape, type, and volume fraction ($\phi$) (see Fig. 1) to achieve the desired thermal resistance, dielectric isolation, moisture resistance, and breakdown voltage.

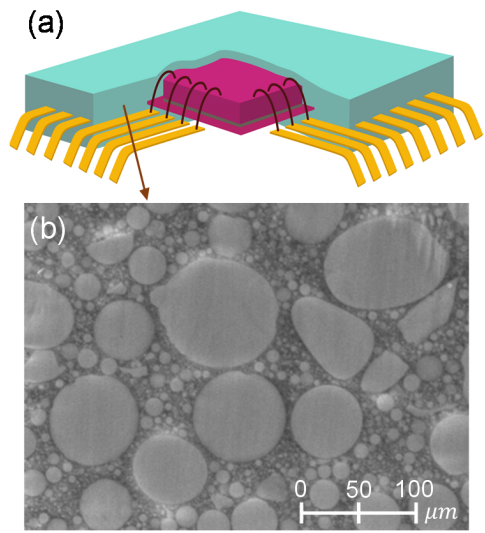

**Figure 1: (a)** 3D schematic of Quad Flat Packaged device encapsulated with composite material. **(b)** SEM image of cross-section view of a commercial micro-composite material.

Moisture uptake in bulk polymers was often described by Fickian diffusion models originally developed for inorganic materials. However, a variety of experiments demonstrated that the Fickian model cannot predict the anomalous mass-gain due to the reaction between polymer chains and water molecules [4, 5]. Since then, a bulk-Langmuir model has been used to describe reaction-diffusion of water within the polymer, where water molecules can reside either in the mobile phase ($w$) or in the reacted phase ($Y$) [6, 7]. This distinction is important because ion diffusion depends exclusively on $Y$ concentration. A generalization of the bulk-Langmuir model that describes the *filler-fraction dependent moisture transport* in MC is necessary to describe ion-transport in these composite materials. Such a model has neither been developed nor experimentally validated.

In this work, we (i) generalize the Langmuir model for moisture uptake in MCs by explicitly accounting for the filler properties and propose generalize effective medium and solubility (*GEMS*) Langmuir model; (ii) validate the *GEMS*- Langmuir model theoretically by finite element method (FEM) simulation [8] and empirically by water uptake experiments for various MCs with different filler configurations; (iii) reassess extracted parameters to successfully estimate $Y$ amount of any arbitrary structure of MCs; (iv) show $Y$ plays a dominant role on $\sigma$ change by experiments; and finally, (v) investigate the impact of filler configuration (e.g., fillers amount, fillers size distribution, and adhesion promoter existence) on HV-ICs application MCs. The *GEMS*-Langmuir can be adopted to describe more detailed anomalous water diffusion, suggesting new opportunities for co-optimization and reliability prediction for traditional and emerging MCs.

## 2. Materials and experiments

### 2.1. Water uptake measurement

Gravimetric measurements have been used to quantify the moisture absorption kinetics within the test structures. The samples include a variety of epoxy-based MCs loaded with different concentrations of silica micro-fillers, see Table 1. In this work, the fill-fraction ($\phi_f$) varies from 60 to 84%. Although, random filling with identical spheres cannot exceed 64%, the polydisperse-size distribution of the fillers allows $\phi_f$ to reach up to 84% [9].

The moisture absorption of the EMCs has been investigated by using a climatic chamber (Genviro-060-C) to control the temperature and the relative humidity. Before moisturizing treatments, the samples have been dried at 125 °C for 24 hours to remove any residual humidity and weighted with an analytical balance (Sartorius CP 124 S) with an accuracy of 0.0001 g to determine their dry weight M0. Subsequently, the samples have been placed into the climatic chamber at 85 °C with two different RH conditions (i.e., 60 % and 85 %) and periodically weighted until equilibrium was reached. The measurements follow the test protocol specified in [10]. The results are summarized in Figures 2 and 3.

**Table 1:** Filler configuration and geometric information of MCs used in this work for mass gain experiments.

|  | **M73** | **M90** | **M91** | **MX87** (No adhesion Promoters) |
|---|---|---|---|---|
| Mass [$g$] | 4.88 | 5.54 | 5.72 | 5.66 |
| Size [$mm^3$] | 59×64×0.75 | 59×64×0.75 | 59×64×0.75 | $\pi 22.5^2 \times 2$ |
| Filler amount [%] | 73 / 60 (Weight/Volume) | 90 / 83 (Weight/Volume) | 91 / 84 (Weight/Volume) | 87 / 74 (Weight/Volume) |
| Filler size [$\mu m$] | ~20 / ~75 (Average/Max) | ~25 / ~135 (Average/Max) | ~20 / ~75 (Average/Max) | ~20 / ~75 (Average/Max) |

## 2.2. DC current measurement

To measure the steady state conductivity, a 3 kV DC step voltage (or, equivalently an electric field of 4 kV/mm) was applied across a sample in a thermostatic oven and the resulting current was monitored with a Keithley-6514 electrometer for 1000 seconds. The measurement time is long enough to separate polarization effects and carrier conduction, but short enough to avoid any significant moisture desorption. The conductivity has been extracted from the steady-state current $I$ as: $\sigma = Ih/(A \cdot V)$, where $h$, $V$, and $A$ are the thickness of the sample, the applied voltage, and the area of the electrodes, respectively. The experimental setup allows to measure a minimum conductivity of about $10^{-19}$ S/cm. The analysis has been carried out on a sample of each material both in dry and under humidity in saturation conditions at the temperature of 25 °C. The conductivity vs. fill fraction is plotted in Fig. 4.

## 3. Analytical model derivation

Langmuir model is widely used because it describes the quasi-saturation of the gravimetric curves [4, 5]. Indeed, many experiments have demonstrated a two-stage water uptakes behavior and many authors proposed several hypotheses [11, 12] regarding the specific reactions involved. Although the hypotheses are not the same, however, there is a consensus that anomalous water diffusion must be described based on a diffusion-reaction scheme. In this case, the time-dependent water uptake is given by Eq. 1 [6].

$$\frac{M_t}{M_\infty} = 1 - \frac{k_F}{k_F + k_R}e^{-k_R t} - \frac{8k_R}{(k_F + k_R)\pi^2}\sum_{n=0}^{\infty}\frac{1}{(2n+1)^2}e^{\frac{-D_{eff}(2n+1)^2\pi^2}{4L^2}t} \quad (1)$$

where $k_F$, and $k_R$ are forward and reverse reaction rates for the chemical reaction (i.e., $w \rightleftarrows Y$), respectively. $M_t$ is $(W_t - W_0)/W_0$. $W_t$, and $W_0$ are the sample weight after time $t$, and initial sample weight, respectively. Thus, $M_t$, $M_\infty$ corresponds to the ratio of absorbed water weight to initial weight at time $t$, and $(W_\infty - W_0)/W_0$, respectively. Once the model constants $k_F, k_R, D_{eff}$, and $M_\infty$ are obtained from calibration experiments of a polymer, standard Langmuir model can be used predict its time-response to variety of moisture concentrations. This simple model does not apply directly for MC with high fill-fraction (and must therefore be generalized) for the following reasons.

### 3.1. Proposed Generalized Effective Medium & Solubility (GEMS) Langmuir model

If the chemical components composition of the polymer matrix (e.g., hardener, adhesion promoters, and other additives) is identical, then $k_F$ and $k_R$ would not change. Among other parameters, the amount of filler affects $D_{eff}$ in MCs, environmental conditions affect the value of $M_\infty$, as discussed below. As a result, the standard Langmuir model will require testing of large number of samples with different fill fraction and tabulating the coefficients. This empirical approach is clearly unsatisfactory. Instead, if we could develop a theory for $D_{eff}(\phi_f)$, we will be able to extract all the parameters from a single sample of a given fill-fraction and then predict transient water uptake behavior at any other fill-fraction and RH in the environment.

#### 3.1.1. Theory of Fill-fraction dependent Diffusion coefficient, $D_{eff}(\phi_f)$

For a composite material, effective physical properties can be calculated by the generalized effective medium theory (EMT) model [13, 14]. In [15], the $D$ at the interface between filler and mold is reported to be 1.5~5 times higher than the bulk $D$. Therefore, for smaller fillers, as in nanocomposites, a three-component treatment (i.e. bulk, interface, and filler) may be necessary [3, 16]. For MCs with large fillers (> 20 um) considered in this paper and for high-contrast system (e.g., $D$ of water in polymer $\gg$ in silica filler), the MC can be approximated as a two-component (i.e. bulk and filler) system. Moreover for high-contrast systems, effective properties depend primarily on network connectivity, as in the case of percolation theory [17]. For these systems, McLachlan developed generalized effective medium (GEM) theory by introducing the percolation threshold and scaling exponents into the symmetric EMT equation as shown in Eq. 2 [18].

$$\phi_f \frac{D_f^{\frac{1}{t}} - D_{\text{eff}}^{\frac{1}{t}}}{D_f^{\frac{1}{t}} + (p_c^{-1} - 1)D_{\text{eff}}^{\frac{1}{t}}} + (1 - \phi_f) \frac{D_m^{\frac{1}{s}} - D_{\text{eff}}^{\frac{1}{s}}}{D_m^{\frac{1}{s}} + (p_c^{-1} - 1)D_{\text{eff}}^{\frac{1}{s}}} = 0 \qquad (2)$$

where $D_m$, $D_f$, $\phi_f$, $p_c$, $s$, and $t$ are water diffusivity in the mold, in the fillers, filler volume fraction, void percolation threshold (from mold perspective), and empirical fitting parameters, respectively. Although the model is empirical, its predictions have been validated both experimentally and numerically [19]. Since water cannot penetrate the fillers, we set $D_f$ to '0'. In this special case and for low-to-moderate fill fraction (e.g., $\phi_f \sim 0$-60%), Eq. 2 is numerically identical to modified Maxwell-Garnett (MG) model described in [20]. As $\phi_f$ approaches to '$1 - p_c$', the curve deviates from modified MG model [20] and eventually, $D_{\text{eff}}$ reduces to zero at '$1 - p_c$'. Since $\phi_f$ of typical mold compound is high, a precise parameterization of Eq. 2 (i.e., $s$, $t$, and $p_c$) are necessary for an accurate prediction.

### 3.1.2. Relationship between Bulk vs. Surface Concentration

Once the $D_{\text{eff}}(\phi_f)$ is known from Eq. (2), we need to determine time-dependent water ingress in the mold compound if we can relate the RH and water concentration at the interface. The relationship between RH and the maximum water concentration ($w_{\text{Max}}$) in a mold can be determined by Henry's law ($w = S^* \cdot p_w$, where $S^*$ and $p_w$ are solubility of water in the polymer, and partial pressure in the environment, respectively). Once we know $w_{\text{Max}}$, $M_\infty$ can be calculated by multiplying water molar mass ($18[g/mol]$), and volume of the MC [$cm^3$] as shown in Eq. 3.

$$p_w = a_w \cdot e^{13.756 - \frac{m}{T}} \tag{3a}$$

$$W_\infty = S^* \cdot p_w \cdot 18 \cdot V \tag{3b}$$

$$S^* = S^{\text{ref}} \cdot \frac{1 - \phi_f^*}{1 - \phi_f^{\text{ref}}} \tag{3c}$$

$p_w$ [$Pa$] can be expressed by assuming the vapor pressure of water satisfies Rankine's formula as shown in Eq. 3a (i.e., general equation), where $m$ (5120 [8]) is specific constant for each substance. Water activity factor ($a_w$) in the environment can be defined as 0.4, 0.6, and 0.9 for RH of 40%, 60%, and 85%, respectively, based on the observation in the literature that shows $a_w$ is always smaller than unity [21], and increases with as RH increases. In this work, $S^*$ [$mol \cdot m^{-3} \cdot Pa^{-1}$] is assumed to have a linear relationship with $\phi_m$ for a given unit volume of MC as shown in Eq. 3c. Once we obtain $S^{\text{ref}}$ from the experiment data of the given sample ($\phi_f^{\text{ref}}$), we can estimate $S^*$ of the any given MC ($\phi_f^*$) with arbitrary $\phi_f$.

Finally, $D_{\text{eff}}$ and $M_\infty$ in Eq. 1 can be revised to Eq. 2, and Eq. 3b ($M_\infty = (W_\infty - W_0)/W_0$) that enclose the information of MCs filler configuration and RH of the environment. In this paper, we applied these revised parameters in conventional Langmuir model, and refer to *GEMS*-Langmuir model. In the following section, the analytical *GEMS*-Langmuir model is validated by detailed FEM simulation.

## 3.2. Numerical (FEM) Validation of *GEMS* Langmuir model

Fig. 2 compares the predictions by 2D FEM simulation[8] and *GEMS*-Langmuir model. Since $D_{\text{eff}}$ reduces with $\phi_f$, therefore the water uptake is reduced if $\phi_f$ is increased (for a fixed RH=85%), as shown in Fig. 2a, top. Also, higher RH increases both the surface concentration and bulk water uptake, see Fig. 2b, bottom. Since the analytical *GEMS*-Langmuir model reproduces the FEM simulation results for arbitrary fill-fraction and RH, we can extract all the relevant parameters from a single experiment and predict transient water uptake at arbitrary conditions. We validate this assertion in the next section.

## 3.3. Experimental Validation of *GEMS* Langmuir model

The goal of this section is to show a single-sample analysis based on Eqs. (2) and (3) produces the equivalent information as many-sample analysis based on Eq. (1). Among the parameters of the Eq. 1, the four parameters ($k_F$, $k_R$, $D_{\text{eff}}(\phi_f)$, and $S^*$, included in $M_\infty$) are obtained by fitting the experimental data for *each* sample (with known $\phi_f$), while the remaining three parameters ($W_0$, $V$, and $p_w$) are known a-prior. All experiments are done at 85°C and 85% of RH as shown in Fig. 3a. The fitting parameters are extracted by a nonlinear least squares regression curve fitting tool, available in MATLAB [22]. These empirically fitted parameters are then compared with the predictions of *GEMS*-Langmuir model equations (i.e., Eq. 2, and Eq. 3b) to validate the model. Fig. 3 shows that GEMS-Langmuir model anticipates very well the parameters obtained by fitting the experimental data from many samples. Thus validated, GEMS-Langmuir model obviates the need for testing multiple samples. The fitting of the measured data by the generalized theory allows us to carefully explore the functional dependencies observed in the experiments, as follows.

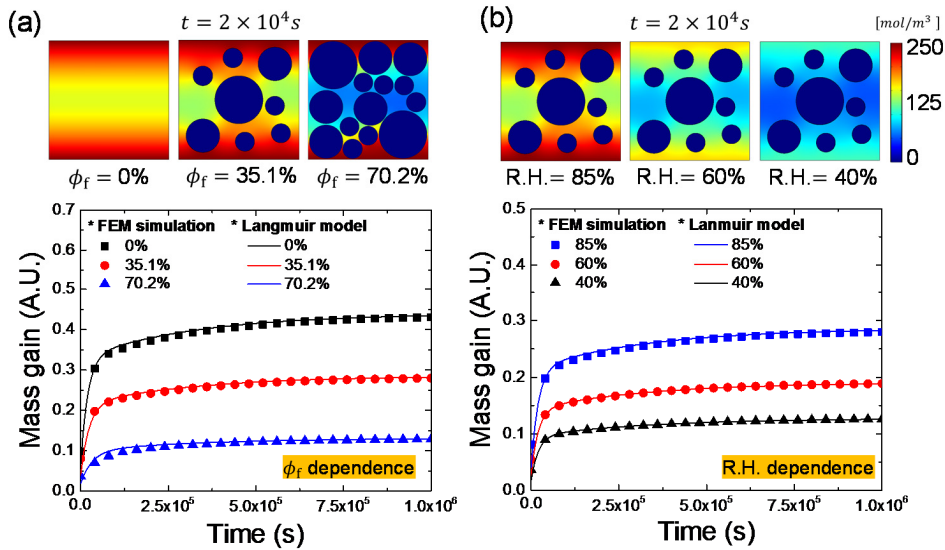

**Figure 2:** Water ($w$) concentration comparison profile comparison for different $\phi_f$, and RH at $t$=20000s. Detailed FEM simulation and analytic gravimetric curve based on *GEMS*-Langmuir model comparison for different **(a)** $\phi_f$, and **(b)** R.H. ($\phi_f$=35.1% case)

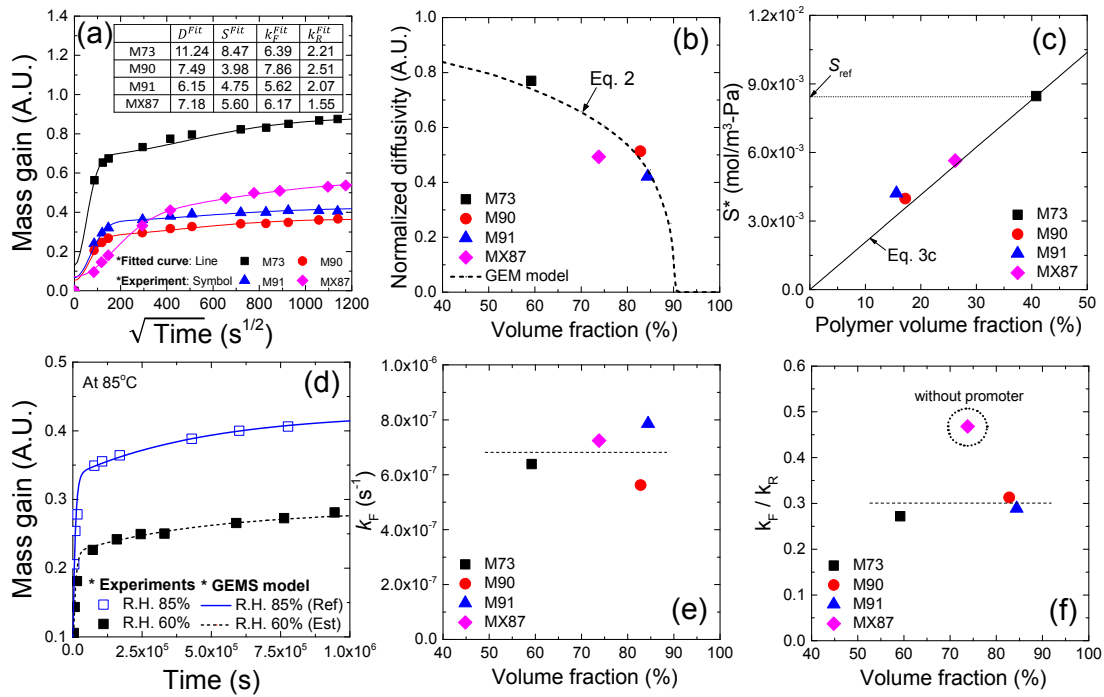

**Figure 3:** **(a)** water uptake experiment and fitted parameters of all MCs. **(b)** Extracted $D$ exhibits clear agreement with Eq. 2, where the value of $s$, $t$, and $p_c$ are 0.3, 1, and 0.9, respectively. The correlation is similar to the work shown in [8-10]. **(c)** Validation of linear correlation between $S$ and $\phi_m$. **(d)** Experimental validation of *GEMS*-Langmuir model when RH is varied. **(e)** Extracted $k_F$ exhibits similar for all MCs. **(f)** $k_F/k_R$ for M73, M90, and M91 are similar, but differs from the sample without adhesion promoter (M87).

### 3.3.1. Filler-size vs. volume fraction dependent moisture diffusion (Eq. 3b and Fig. 3b)

The experimental results plotted in Fig. 3b show that the $D_{\text{eff}}$ decreases with $\phi_{\text{f}}$, but an explicit functional dependence is not obvious. GEMS-Langmuir model, however, can not only fit that data, but also predict the fill-fraction necessary to suppress water diffusion [18, 19].. Interestingly, Table 1 shows that the samples M90 and M91 have similar $\phi_f$, but very different size distribution. The similarity of the $D_{\text{eff}}$ of these two samples confirms that fill-fraction (not the size distribution) determine $D_{\text{eff}}$, consistent with the prediction of the GEMS-Langmuir model.

### 3.3.2. RH and volume-fraction dependent saturation density (Eq. 3c and Fig. 3c)

GEMS-Langmuir model accurately describes the volume-fraction dependent $S(\phi_f)$. Unlike $D(\phi_f)$, $S(\phi_f)$ defines the local maximum available water amount, therefore it does not distort the concentration field in a composite material. Fig. 3c shows that the linear relation between $\phi_{\text{m}}$ and $S^*$ anticipated by Eq. 3c is validated by experimental results.

### 3.3.3. Experimental validation of time-kinetics predicted by Eqs. (2)-(3)

Once we have a model for $D(\phi_f)$ and $S(\phi_f)$ through Eq. 3, the GEMS-Langmuir model should predict the time-kinetics at arbitrary $\phi_f$ and RH. Fig. 3c shows that this is indeed the case. Water uptake experiment under different RH is also done. Based on extracted $k_{\text{F}}$, $k_{\text{R}}$, $D$, and $S^*$ of M90 at 85°C and 85% of RH can be applied to predict mass uptake behavior at different RH. Indeed, Fig. 3d shows that the time-kinetics for RH=60% is predicted by the theory.

### 3.3.4. Effect of adhesion promotor

The two-component GEMS-Langmuir theory does not account for adhesion promoter explicitly, therefore any deviation from the model prediction may be attributed to adhesion promoters. For example, the model predicts that $k_F$ and $k_F/k_R$ should be independent of $\phi_f$. Experiments in Fig. 3e and 3f shows that this is true for $k_\mathrm{F}$, but not for $k_\mathrm{F}/k_\mathrm{R}$. The outlier is MX87, samples *without* adhesion promoter. This poor adhesion enhances $Y$ generation, explaining the higher $k_\mathrm{F}/k_R$ compared to other samples.

To conclude this section, GEMS-Langmuir theory predicts water uptake behavior of other MCs with different $\phi_\mathrm{f}$, or under different RH. In the following section, we investigate the optimization of filler configuration for reliable HV-ICs encapsulant based on parameters extracted based on *GEMS*-Langmuir model, and electric-field distribution simulation.

4. **Results and Discussions: Effect of RH on conductivity and dielectric breakdown**

For high reliability, the HV-IC encapsulants must have low electrical conductivity ($\sigma$) and high dielectric strength ($E_{BD}$) to sustain high voltage applied during operation. Lower $\sigma$ reduces charge injection from electrodes [23], suppress charges migration [24], and minimize energy dissipation in AC circuit [25]. In the following section, we will discuss how filler configuration and RH impact $E_{BD}$ and the dissipation factor, $\sigma/(\omega \cdot \varepsilon_\mathrm{eff})$, where $\omega$, and $\varepsilon_\mathrm{eff}$ are operating frequency multiplied by $2\pi$ and dielectric constant, respectively.

4.1. **Moisture-dependent changes in the mold electrical conductivity**

Although many experiments have focused on water uptake epoxy MCs, the role of water in defining $\sigma$ is not fully understood [26]. A recent work shows that hydrolysis, and the reaction between moisture and anhydride groups ($Y$) increase $\sigma(Y)$ significantly, whereas the effect of mobile water ($w$) is negligible [27]. In this work, we calculate $Y$ concentration analytically and correlate the concentration to the conductivity measurement. Recall that.

$$w + Y = \frac{W_\infty - W_0}{18 \cdot V} \tag{4a}$$

$$Y = (w + Y) \cdot \left(\frac{k_F/k_R}{1 + k_F/k_R}\right) \tag{4b}$$

$$Y_m = \frac{Y}{1 - \phi_f} \tag{4c}$$

where $W_\infty - W_0$ [$g$], 18 [$g/mol$], and $V[m^3]$ are mass gain due to water, water molar mass, and sample volume, respectively. From mass gain experiments, we can calculate the sum of reacted and free water concentration (i.e., $w + Y[mol/m^3]$) within the MC, see Eq. 4a. Based on Eq. A3 ($Y = w \cdot (k_F/k_R)$, at steady state), $Y$ concentration is obtained by multiplying the term in the second parenthesis of Eq. 4b. Finally, Eq. 4c determines $Y$ within mold ($Y_m$) by normalizing with $(1 - \phi_f)$, since silica fillers are impermeable to water molecules. This explains why MX87 has the highest $Y_m$, even though M73 has the highest water gain (Fig. 3a.)

### 4.1.1. Effect of adhesion promoter on $\sigma$

Fig. 4b shows that while M73 show the highest $Y_m$, the sample without adhesion promoter (MX87) actually has higher $\sigma$. Fig. 4 also shows that $Y_m$ calculated from theory and wet $\sigma$ measured from the experiments are correlated. Therefore, we conclude that adhesion promoter is essential to lower $\sigma(Y_m)$ by reducing available reactive sites, and $Y$ byproduct after water uptake in a polymer-based encapsulant.

### 4.1.2. Effect of filler volume fraction on $\sigma$

In sec. II, we observed the impact of $\phi_f$ on $D$, and $S$. Thus, $Y$ increasing over time before characteristic time (i.e., $0.67 \cdot L^2/D$, the time when water saturates the sample) also depends on $\phi_f$. In a longer time scale, however, $\phi_f$ impact on $Y_m$ content is actually negligible as shown in Fig. 4a. In other words, although the total water content increases when $\phi_f$ is reduced, but $Y_m$ are nominally independent of $\phi_f$. Thus, $\phi_f$-optimization should focus on other properties (e.g., higher thermal conductivity, higher stiffness, and lower dielectric constant) rather than $\sigma$ in a humid environment.

### 4.2. Moisture-dependent changes in the mold dielectric constant

Our previous experiments [28, 29] have shown that an effective dielectric constant ($\varepsilon_{\text{eff}}$) of MC can be predicted by an appropriately generalized EMT model. There are two widely known EMT models: MG model [20], and Bruggeman model [30] as shown in Eq. 5.

$$\varepsilon_{\text{eff}} = \varepsilon_m \left[ 1 + \frac{3(\varepsilon_f - \varepsilon_m)\phi_f}{\varepsilon_f + 2\varepsilon_m - (\varepsilon_f - \varepsilon_m)\phi_f} \right] \tag{5a}$$

$$\phi_f \frac{\varepsilon_f - \varepsilon_{\text{eff}}}{\varepsilon_f + 2\varepsilon_{\text{eff}}} + (1 - \phi_f) \frac{\varepsilon_m - \varepsilon_{\text{eff}}}{\varepsilon_m + 2\varepsilon_{\text{eff}}} = 0 \tag{5b}$$

where $\varepsilon_f$, and $\varepsilon_m$ are dielectric constant of filler, and mold, respectively. The MG model calculates the filler polarization induced by the external field, and presumes the dipole on one quasi-spherical filler does not affect its neighbors. Thus, the model is accurate when $\phi_f$ is low. On the contrary, Bruggeman model accounts for the calculation at high $\phi_f$, namely, aggregate structure. We note that Eq. 5b can be derived from Eq. 2 by putting 1, 1, and 1/3 into $s$, $t$, and $p_c$, respectively. Thus, Bruggeman model is appropriate for low-contrast systems where percolation behavior is insignificant. The difference between MG, and Bruggeman model becomes considerable when $\varepsilon_f$ differs significantly from $\varepsilon_m$. In the case of typical commercial MC with $SiO_2$ fillers, the dielectric contrast is small for both dry and wet conditions. Thus, the two models result in almost identical curves as shown in Fig. 5. Therefore, we can estimate $\varepsilon_{\text{eff}}$ of a MC either by the MG model or the Bruggeman

model.where $\varepsilon_f$, and $\varepsilon_m$ are dielectric constant of filler, and mold, respectively. The MG model calculates the filler polarization induced by the external field, and presumes the dipole on one quasi-spherical filler does not affect its neighbors. Thus, the model is accurate when $\phi_f$ is low. On the contrary, Bruggeman model accounts for the calculation at high $\phi_f$, namely, aggregate structure. We note that Eq. 6b can be derived from Eq. 3 by putting 1, 1, and 1/3 into $s$, $t$, and $p_c$, respectively. Thus, Bruggeman model is appropriate for low-contrast systems where percolation behavior is insignificant. The difference between MG, and Bruggeman model becomes considerable when $\varepsilon_f$ differs significantly from $\varepsilon_m$. In the case of typical commercial MC with $SiO_2$ fillers, the dielectric contrast is small for both dry and wet conditions. Thus, the two models result in almost identical curves as shown in Fig. 7. Therefore, we can estimate $\varepsilon_{eff}$ of a MC either by the MG model or the Bruggeman model.

### 4.3. Moisture-dependent dielectric strength of mold compounds ($E_{MC}^{Crit}$)

In order to study the impact of filler configuration on dielectric strength of MC, ($E_{MC}^{Crit}$ i.e., $V/L$), we compare the local field distribution within each sample by FEM simulation.

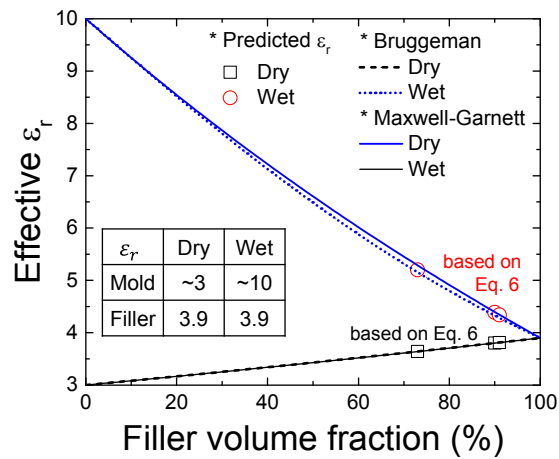

**Figure 5:** Calculated $\varepsilon_{eff}$ based on EMT model, validated in our previous work [57]. The difference between MG model, and Bruggeman model is insignificant due to small contrast ratio between $\varepsilon_f$, and $\varepsilon_m$. Thus, both model can be applied to estimate $\varepsilon_{eff}$ of MCs.

The dielectric strength is an intrinsic property of a material (e.g., dielectric strength of SiO$_2$ ($E_{\text{f}}^{\text{Crit}}$), and epoxy ($E_{\text{m}}^{\text{Crit}}$) are ~5MV/cm [31], and ~0.5MV/cm [32], respectively). If the local maximum electric field inside the epoxy ($E_{\text{Max}}$) exceeds $E_{\text{MC}}^{\text{Crit}}$, the MC will break [33]. The $E_{\text{Max}}$ will vary depending on the filler configuration so that comparison of electric field cumulative distribution function around $E_{\text{Max}}$ ($F(E)$) is crucial to evaluate $E_{\text{MC}}^{\text{Crit}}$ for each MC with different fillers configuration, or different surrounding environment. For example, if $F(E)$ of sample A exhibits less $E_{\text{Max}}$ (i.e., electric field strength at the cumulative percentage of 100%) than that of sample B, $E_{\text{MC}}^{\text{Crit}}$ of sample A will be stronger. Simulated structures are constructed based on the actual configuration (e.g., $\phi_{\text{f}}$, and the average diameter of the fillers) to emulate real samples as shown in Fig. 6a. This simplified geometry can be also used for the actual case studies based on the fact that nearest neighbor function between an actual and a simplified structure exhibits similarly [34].

### 4.3.1. Adhesion promoter effect

When the sample is dried, $\varepsilon_{\text{m}}$, and $\varepsilon_{\text{f}}$ are ~3.0, and 3.9, respectively, so that the ratio of $\varepsilon_{\text{f}}$ to $\varepsilon_{\text{m}}$ is greater than 1. However, when the sample is exposed to a humid environment, considering $\varepsilon_{\text{r}}$ of water is around 80 at room temperature, the ratio will be smaller than 1. As more water is absorbed into the sample, smaller the ratio will be. Electric field distributions for the different ratio is shown in Fig. 6b. All situations exhibit two main peaks in the electric field distribution [35]. The left and the right side peaks correspond to the region with higher $\varepsilon_{\text{r}}$ and lower $\varepsilon_{\text{r}}$ regions, respectively. As water content builds within the sample, we observe that the main peak gradually shifts to the right and the cumulative tail around $E_{\text{Max}}$ gradually extends. Consequently, the water content ($w + Y$) increases $E_{\text{Max}}$ of the sample so that $E_{\text{MC}}^{\text{Crit}}$ will become weaker. In other words, reducing $Y$ by adding adhesion promoter will further enhances $E_{MC}^{Crit}$.

### 4.3.2. Filler configuration (volume fraction, and size distribution) effect

When the sample is dried, the experiments show a decrease in $E_{MC}^{Crit}$ with increasing $\phi_f$ [36]. Similarly, our simulations show that $E_{Max}$ of M90, and M91 is higher than that of M73, see Fig. 6c. However, once $\varepsilon_m$ becomes larger than that of $\varepsilon_f$ due to water ingress, $E_{Max}$ of M73 becomes higher than that of M90, and M91 as shown in Fig. 6d. Compared to the dry samples, electric field distribution under the wet condition is reversed as shown in Fig. 6a

Even if $\phi_f$ is the same, electric field distribution can be different if the size distribution is dissimilar. When the average radius is larger (e.g., M90), the distribution for the main peak in the filler-phase exhibits narrower peak. Based on the nearest surface distribution function, we realize that bigger particles have more probability of having more neighbors a shorter distance away [37], resulting in a concentrated distribution due to the filler phase. However, the tail of the right side distribution which determines maximum electric field does not change significantly with different filler-size distribution. Therefore, $\phi_f$ should be carefully optimized depending on the severity of humid environment: in a dry environment, lower $\phi_f$ samples are preferred, whereas, in a humid environment, higher $\phi_f$ samples would be more reliable.

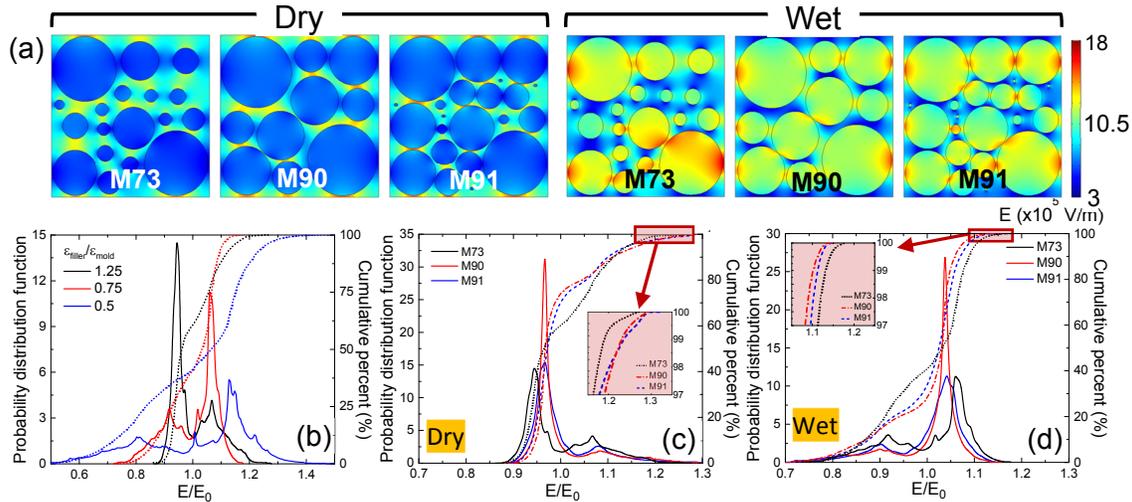

**Figure 6: (a)** 2D FEM simulation showing electric field distribution. Comparison of electric field cumulative distribution depends on **(b)** different water amount in polymer **(c)** different MCs at dry condition **(d)** different MCs at wet condition. In dry case, lower $\phi_f$ (M73) exhibits the strongest $E_{MC}^{Crit}$. However, M73 shows the weaker $E_{MC}^{Crit}$ when the MC absorbs moisture.

## 5. Conclusions

Among many encapsulant properties, achieving lower dissipation factor, and higher $E_{MC}^{Crit}$ are important for HV-ICs applications. In this paper, we show *GEMS*-Langmuir model can predict the water uptake behavior for MCs with any filler configuration under a variety of environmental conditions. We also compared electric field distribution for each MC in both dry and wet environment. Based on the *GEMS*-Langmuir model and experiments, following conclusions can be drawn.

- $Y$ (rather than $w$) plays a dominant role in determining $\sigma$ of MCs.
- Adhesion promoter is essential for lower $\sigma$, and higher $E_{MC}^{Crit}$ in humid environment.
- $\phi_f$ does not impact on $\sigma$ significantly, however, it needs to be carefully optimized for higher $E_{MC}^{Crit}$ depends on the severity of humidity.

Once all the parameters are known from mass gain experiments, *GEMS*-Langmuir model provides a deep and fundamental new insight regarding the physics of moisture transport in MCs. The model will also be useful initial optimization of MCs for other applications involving polymer encapsulants, e.g., solar cells and biosensors.

## Appendix

We validated *GEMS*-Langmuir model by the following FEM simulation set up introduced in [19] for water absorption. More details can be found in [19]. Conservation of water ($w$), reactive sites ($R$), and reacted water ($Y$) in the system is shown in Eq. A1.

$$\frac{\partial w}{\partial t} = -\nabla \cdot J + r_w, \quad \frac{\partial R}{\partial t} = r_R, \quad \frac{\partial Y}{\partial t} = -r_Y \tag{A1}$$

where $J$, and $r$ are the water flux, and reaction rate, respectively. The water flux derived from the chemical potential of diffusing species is shown in Eq. A2.

$$J = -D\left(\nabla w - w \cdot \frac{\nabla(S_0 + S_1 \cdot Y) \cdot p_w}{(S_0 + S_1 \cdot Y) \cdot p_w}\right) \tag{A2}$$

where $S_0$, and $S_1$ are the initial solubility of water in mold, and optimization parameter to include the effect of $Y$ on solubility. Each reaction term is shown in Eq. A3

$$-k_h \cdot w \cdot R + k_R \cdot Y = r_w = r_R = -r_Y \tag{A3}$$

One of the fitting parameters in *GEMS*-Langmuir model ($k_F$) is same as $k_h \cdot R$ in Eq. A3. In this work, The boundary conditions are set to $(S_0 + S_1 \cdot Y) \cdot p_w$. Finally, the initial condition is shown in Eq. A4.

$$w = 0, \ R = R_0, \ Y = 0 \text{ at } t = 0 \tag{A4}$$